Relativistic quark model and lowest hybrid mesons.

Gerasyuta S.M.*, Kochkin V.I.

Department of Theoretical Physics, St. Petersburg State University, 198904, St. Petersburg, Russia.

Abstract.

The relativistic four-quark equations are found in the framework of the dispersion relation technique. The solutions of these equations using the method based on the extraction of leading singularities of the amplitudes are obtained. The mass spectrum values of lowest hybrid mesons are calculated.

---

*  Present address: Department of Physics, LTA, Institutski Per. 5, St. Petersburg 194021, Russia

## I. Introduction.

The present understanding of strong interactions is that they are described by QCD. This non-Abelian field theory not only describes how quarks and antiquarks interact but also predicts that the gluons which are the quanta of the field will themselves interact to form mesons. If the object formed is composed entirely of constituent gluons ($gg$) the meson is called a glueball, however if it is composed of a mixture of constituent quarks, antiquarks and gluons ($q\bar{q}g$) it is called hybrid. In addition, $q\bar{q}q\bar{q}$ states are also predicted. An unambiguous confirmation of these states would be an important test of QCD and give fundamental information on the behavior of this theory in the confinement region.

Historically, there have been two approaches to the consideration of hybrids. The first assumes that hybrids are predominantly quark-antiquark states with an additional constituent gluon [1 - 4] and that decays proceed via constituent gluon dissociation [5 - 7]. The second assumes that hybrids are predominantly quark-antiquark states moving on an adiabatic surface generated by an exited "flux tube" configuration of glue [8]. Decays then proceed by a phenomenological pair production mechanism (the "$^3P_0$ model" [9 - 13]) coupled with a flux tube overlap [14, 15].

In the recent papers [16, 17] the relativistic four-quark equations are represented in the form of the dispersion relation over the two-body subenergy. The behavior of the low-energy four-particle amplitude is determined by its leading singularities in the pair invariant masses. The suggested method of approximate solutions of the relativistic four-quark equations was verified on the example of the lowest cryptoexotic mesons [18]. The calculated mass values of cryptoexotic mesons are in good agreement with the experimental ones.

In the present paper we calculated the masses of the lowest hybrid mesons using the method based on the extraction of leading singularities of the amplitude. The interesting result of this model is the calculation of hybrid meson amplitudes, which contain the contributions of two subamplitudes: four-quark amplitude and hybrid amplitude. The contributions of these subamplitudes with different quantum numbers are given in Table I. One can see that the main contribution corresponds to the four-quark amplitude. The hybrid amplitude give rise to only less 40 % of the hybrid meson contributions.

In Section II the relativistic four-quark equations are constructed in the form of the dispersion relation over the two-body subenergy. The approximate solutions of these equations using the method based on the extraction of leading singularities of the amplitude are obtained. The quark amplitudes of hybrid mesons are calculated.

Section III is devoted to the calculation results for the lowest hybrid meson mass spectrum (Table I).

In the Conclusion the status of the considered model is discussed.

In the Appendix A the quark-antiquark vertex functions and the phase spaces for the hybrid mesons are given (Tables II, III) respectively.

In the Appendix B we search the integration contours of functions $J_1$, $J_2$, $J_3$, which are determined by the interaction of the four quarks.

## II. Quark amplitudes of the hybrid mesons.

In the present paper we investigate scattering amplitudes of the constituent quarks of two flavours (u, d). The poles of these amplitudes determined the masses of the lowest hybrid mesons. The constituent quark is color triplet and quark amplitudes obey the global color symmetry. One uses the results of the bootstrap quark model [19, 20] and introduce the $q\bar{q}$ amplitude in the colour octet channel $c = 8_c$ with $J^{PC} = 1^{--}$. This bound state should be identified as constituent gluon with mass of the order of 0.7 GeV. In our consideration we take into account the colour octet state with $J^{PC} = 1^{--}$ and isospin I=1, which determines with the constituent gluon the hybrid state $q\bar{q}g$. Now we have the mixing between hybrid and $q\bar{q}q\bar{q}$ states. This state is called the hybrid meson.

We derived the relativistic four-quark equations in the framework of the dispersion relation technique. Let some current produce two pairs of quark-antiquark (Fig.1). The diagrams in Fig.1 allows to graphically present the equations for the four-quark amplitudes. However, the correct equations for the amplitude are obtained at the account of all possible subamplitudes. This corresponds to split complete system into subsystems from the smaller number of particles. Then one should present four-particle amplitude as a sum of six subamplitudes: $A = A_{12} + A_{13} + A_{14} + A_{23} + A_{24} + A_{34}$. This defines the division of the diagrams into groups according to the last interaction of particles. In this case we need to consider only one group of diagrams and the amplitude corresponding to them, for example $A_{12}$. One must take into account each sequence of the inclusion of interaction. For instance, the process beginning with interaction of the particles 1 and 2 can proceed by the three ways: particle 3 and 4 consistently join a chosen pair, or begin to interact among themselves, and each of the three ways of the connection there should correspond to their own amplitudes [21, 22]. Therefore the diagrams corresponding to amplitude $A_{12}$ are divided in three group $A_1(s, s_{12}, s_{123})$, $A_1(s, s_{12}, s_{124})$ and $A_2(s, s_{12}, s_{34})$ (moreover the subamplitudes $A_1(s, s_{12}, s_{123})$ and $A_1(s, s_{12}, s_{124})$ are analogous).

The equations for the four-quark amplitudes in the graphic form are presented (Fig.1). The coefficients are determined by the permutation of quarks [21].

To present the amplitudes $A_1(s, s_{12}, s_{123})$ and $A_2(s, s_{12}, s_{34})$ in form of the dispersion relation it is necessary to define the amplitude of two-quark interaction $a_j(s_{ik})$. One uses the results of the bootstrap quark model [19, 20] and writes down the pair quarks amplitude in the form:

$$a_j(s_{ik}) = \frac{G_j^2(s_{ik})}{1 - B_j(s_{ik})}, \tag{1}$$

$$B_j(s_{ik}) = \int_{m^2}^{\Lambda} \frac{ds'_{ik}}{p} \frac{r_j(s'_{ik})G_j^2(s'_{ik})}{s'_{ik} - s_{ik}}. \tag{2}$$

Here $G_j(s_{ik})$ is the quark-antiquark vertex function. $B_j(s_{ik})$, $r_j(s_{ik})$ are the Chew-Mandelstam function [23] and the phase space respectively. We introduced the cut-off parameter $\Lambda$. There j=1 corresponds to pair of quarks $q\bar{q}$ with isospin I=1 and $J^{PC} = 0^{++}$, $1^{++}$, $2^{++}$, $0^{-+}$, $1^{--}$ (colour singlet $SU(3)_c$) and j=2 defines the quark pair with $J^{PC} = 1^{--}$ in colour channel $8_c$ (constituent gluon). j=3 defines the quark pair with $J^{PC} = 1^{--}$ (isospin I=1) in colour channel $8_c$ (for instance, $u\bar{d}$ state). The vertex functions are shown in the Table II, the functions $r_j(s_{ik})$ are given in the Appendix A (see Table III). In the case in question the interacting quarks do not produce bound state, then the integration in (3) - (4) is carried out from the threshold $4m^2$ to the cut-off $\Lambda$. The integral equation systems, corresponding to Fig 1, have the following form:

$$A_1(s, s_{12}, s_{123}) = \frac{I_1 B_1(s_{12})}{1 - B_1(s_{12})} + \frac{2G_1(s_{12})}{1 - B_1(s_{12})}[\hat{J}_1 A_1(s, s'_{13}, s'_{123}) + \hat{J}_3 A_2(s, s'_{13}, s'_{24})], \tag{3}$$

$$A_2(s, s_{12}, s_{34}) = \frac{I_2 B_2(s_{12}) B_3(s_{34})}{[1 - B_2(s_{12})][1 - B_3(s_{34})]} + \frac{8G_2(s_{12})G_3(s_{34})}{[1 - B_2(s_{12})][1 - B_3(s_{34})]}\hat{J}_2 A_1(s, s'_{13}, s'_{134}). \tag{4}$$

$I_i$ are the current constants. Here we introduce the integral operators:

$$\hat{J}_1(s,m) = \int\limits_{4m^2}^{\Lambda} \frac{ds'_{12}}{\pi} \frac{r_1(s'_{12}) \cdot G_1(s'_{12})}{s'_{12} - s_{12}} \int\limits_{-1}^{+1} \frac{dz_1}{2}, \qquad (5)$$

$$\hat{J}_2(s,m) = \int\limits_{4m^2}^{\Lambda} \frac{ds'_{12}}{\pi} \frac{r_2(s'_{12}) \cdot G_2(s'_{12})}{s'_{12} - s_{12}} \int\limits_{4m^2}^{\Lambda} \frac{ds'_{34}}{\pi} \frac{r_2(s'_{34}) \cdot G_3(s'_{34})}{s'_{34} - s_{34}} \int\limits_{-1}^{+1} \frac{dz_3}{2} \int\limits_{-1}^{+1} \frac{dz_4}{2}, \qquad (6)$$

$$\hat{J}_3(s,m) = \frac{1}{4\pi} \int\limits_{4m^2}^{\Lambda} \frac{ds'_{12}}{\pi} \frac{r_1(s'_{12}) \cdot G_1(s'_{12})}{s'_{12} - s_{12}} \int\limits_{-1}^{+1} \frac{dz_1}{2} \int\limits_{-1}^{+1} dz \int\limits_{z_2^-}^{z_2^+} dz_2 \times$$

$$\times \frac{1}{\sqrt{1 - z^2 - z_1^2 - z_2^2 + 2zz_1z_2}} . \qquad (7)$$

there $m$ are the masses of nonstrange quarks. In the equations (5) and (7) $z_1$ is the cosine of the angle between the relative momentum of the particles 1 and 2 in the intermediate state and that of the particle 3 in the final state, which is taken in the c.m. of particles 1 and 2. In the equation (7) $z$ is the cosine of the angle between the momentum of the particles 3 and 4 in the final state, which is taken in the c.m. of particles 1 and 2. $z_2$ is the cosine of the angle between the relative momentum of particles 1 and 2 in the intermediate state and the momentum of the particle 4 in the final state, which is taken in the c.m. of particles 1 and 2. In the equation (6) we have defined: $z_3$ is the cosine of the angle between relative momentum of particles 1 and 2 in the intermediate state and that of the relative momentum of particles 3 and 4 in the intermediate state, which is taken in the c.m. of particles 1 and 2; $z_4$ is the cosine of the angle between the relative momentum of the particles 3 and 4 in the intermediate state and that of the momentum of the particle 1 in the intermediate state which is taken in the c.m. of particles 3, 4. Using (8) - (12) we can pass from the integration over the cosines of the angles to the integration over the subenergies. The choice of integration contours of functions $J_1$, $J_2$, $J_3$ do not differ from the papers [18] (see Appendix B).

$$s'_{13} = 2m^2 + \frac{s_{123} - s'_{12} - m^2}{2} + \frac{z_1}{2}\sqrt{\frac{s'_{12} - 4m^2}{s'_{12}}[(s_{123} - s'_{12} - m^2)^2 - 4s'_{12}m^2]}, \qquad (8)$$

$$s'_{24} = 2m^2 + \frac{s'_{124} - s'_{12} - m^2}{2} + \frac{z_2}{2}\sqrt{\frac{s'_{12} - 4m^2}{s'_{12}}[(s'_{124} - s'_{12} - m^2)^2 - 4s'_{12}m^2]}, \qquad (9)$$

$$z = \frac{2s'_{12}(s + s'_{12} - s_{123} - s'_{124}) - (s_{123} - s'_{12} - m^2)(s'_{124} - s'_{12} - m^2)}{\sqrt{[(s_{123} - s'_{12} - m^2)^2 - 4m^2 s'_{12}][(s'_{124} - s'_{12} - m^2)^2 - 4m^2 s'_{12}]}}, \qquad (10)$$

$$s'_{134} = m^2 + s'_{34} + \frac{s - s'_{12} - s'_{34}}{2} + \frac{z_3}{2}\sqrt{\frac{s'_{12} - 4m^2}{s'_{12}}[(s - s'_{12} - s'_{34})^2 - 4s'_{12}s'_{34}]}, \qquad (11)$$

$$s'_{13} = 2m^2 + \frac{s'_{134} - s'_{34} - m^2}{2} + \frac{z_4}{2}\sqrt{\frac{s'_{34} - 4m^2}{s'_{34}}[(s'_{134} - s'_{34} - m^2)^2 - 4m^2 s'_{34}]}. \qquad (12)$$

Let us extract two-particle singularities in the amplitudes $A_1(s, s_{12}, s_{123})$ and $A_2(s, s_{12}, s_{34})$:

$$A_1(s, s_{12}, s_{123}) = \frac{a_1(s, s_{12}, s_{123}) B_1(s_{12})}{1 - B_1(s_{12})}, \tag{13}$$

$$A_2(s, s_{12}, s_{34}) = \frac{a_2(s, s_{12}, s_{34}) B_2(s_{12}) B_3(s_{34})}{[1 - B_2(s_{12})][1 - B_3(s_{34})]}. \tag{14}$$

In the amplitude $A_1(s, s_{12}, s_{123})$ we do not extract three-particle singularity, because it is weaker than two-particle and taking into account in the function $a_1(s, s_{12}, s_{123})$.

We used the classification of singularities, which was proposed in papers [17]. The construction of approximate solution of the (13) and (14) is based on the extraction of the leading singularities of the amplitudes. The main singularities in $s_{ik} \approx 4m^2$ are from pair rescattering of the particles i and k. First of all there are threshold square root singularities. Also possible are pole singularities which correspond to the bound states. They are situated on the first sheet of complex $s_{ik}$ plane in case of real bound state and on the second sheet in case of virtual bound state. The diagrams Fig.1 apart from two-particle singularities have their specific triangular singularities and the singularities correspond to the interaction of four particles. Such classification allows us to search the approximate solution of (13) and (14) by taking into account some definite number of leading singularities and neglecting all the weaker ones. We consider the approximation, which corresponds to the single interaction of all four particles (two-particle, triangle and four-particle singularities). The functions $a_1(s, s_{12}, s_{123})$ and $a_2(s, s_{12}, s_{34})$ are the smooth functions of $s_{ik}$, $s_{ijk}$ as compared with the singular part of the amplitudes, hence they can be expanded in a series in the singularity point and only the first term of this series should be employed further. Using this classification one define the functions $a_1(s, s_{12}, s_{123})$ and $a_2(s, s_{12}, s_{34})$ as well as the B-functions in the middle point of the physical region of Dalitz-plot at the point $s_0$:

$$s_0 = \frac{s + 8m^2}{6}, \tag{15}$$

$$s_{123} = 3s_0 - 3m^2. \tag{16}$$

Such a choice of points $s_0$ allows as to replace the integral equations (3) and (4) by the algebraic equations (17) - (18) respectively:

$$a_1 = l_1 + 2a_1 J_1 + 2a_2 J_3 \frac{B_2(s_0) B_3(s_0)}{B_1(s_0)}, \tag{17}$$

$$a_2 = l_2 + 8a_1 J_2 \frac{B_1(s_0)}{B_2(s_0) B_3(s_0)}. \tag{18}$$

Here we introduce following functions:

$$J_1(s, m) = G_1^2 \int_{4m^2}^{\Lambda} \frac{ds'_{12}}{p} \frac{r_1(s'_{12})}{s'_{12} - s_0} \int_{-1}^{+1} \frac{dz_1}{2} \frac{1}{1 - B_1(s'_{13})}, \tag{19}$$

$$J_2(s, m) = G_2^2 G_3^2 \int_{4m^2}^{\Lambda} \frac{ds'_{12}}{p} \frac{r_2(s'_{12})}{s'_{12} - s_0} \int_{4m^2}^{\Lambda} \frac{ds'_{34}}{p} \frac{r_2(s'_{34})}{s'_{34} - s_0} \int_{-1}^{+1} \frac{dz_3}{2} \int_{-1}^{+1} \frac{dz_4}{2} \frac{1}{1 - B_1(s'_{13})}, \tag{20}$$

$$J_3(s,m) = G_1^2 \frac{1-B_1(s_0,\Lambda)}{1-B_1(s_0,\tilde{\Lambda})} \frac{1}{4p} \int_{4m^2}^{\tilde{\Lambda}} \frac{ds'_{12}}{p} \frac{r_1(s'_{12})}{s'_{12}-s_0} \int_{-1}^{+1} \frac{dz_1}{2} \int_{-1}^{+1} dz \int_{z_2^-}^{z_2^+} dz_2 \times$$
$$\times \frac{1}{\sqrt{1-z^2-z_1^2-z_2^2+2zz_1z_2}} \frac{1}{[1-B_2(s'_{13})][1-B_3(s'_{24})]} \quad (21)$$

In our approximation the vertex functions (Table II) are constants. As the integration region the physical region of the reaction should be chosen, therefore $-1 \le z_i \le 1$ ($i=1,2,3,4$). From these conditions we can define the regions of the integration over $s'_{13}$, $s'_{24}$, $s'_{134}$, $s'_{124}$. Let us consider the integration region over $s'_{124}$. For this purpose we use equation (10). This condition corresponds to $0 \le z^2 \le 1$. By consideration of these inequalities one can obtain:

$$s_{124}^{\pm} = s'_{12} + m^2 + \frac{(s-s_{123}-m^2)(s_{123}+s'_{12}-m^2)}{2s_{123}} \pm$$
$$\pm \frac{1}{2s_{123}}\sqrt{[(s_{123}-s'_{12}-m^2)^2 - 4m^2 s'_{12}][(s-s_{123}-m^2)^2 - 4m^2 s_{123}]} \quad (22)$$

We must take into account the upper restriction of the integration region over $s'_{12}$ in $J_3$:

$$\tilde{\Lambda} = \begin{cases} \Lambda, & \text{if } \Lambda \le (\sqrt{s_{123}}+m)^2 \\ (\sqrt{s_{123}}+m)^2, & \text{if } \Lambda > (\sqrt{s_{123}}+m)^2 \end{cases} \quad (23)$$

The integration contours of the functions $J_1$, $J_2$, $J_3$ are given in the Appendix B. The function $J_3$ takes into account the singularity, which corresponds to the simultaneous vanishing of all propagators in the four-particle diagram like those in Fig.1. In the case in question the functions $a_i(s)$ are determined as:

$$a_i(s) = F_i(s, l_i)/\Delta(s) \quad (24)$$

There $\Delta(s)$ is the determinant:

$$\Delta(s) = 1 - 2J_1 - 16J_2 J_3 \quad (25)$$

Right-hand sides of (25) might have a pole in $s$ which corresponds to the bound state of the four quarks. The poles of rescattering amplitudes for the lowest hybrid mesons with $J^{PC} = 0^{++}$, $1^{++}$, $2^{++}$, $0^{-+}$, $1^{--}$ (I=1) correspond to the bound state and determine the masses of the hybrid mesons.

III. Calculation results.

In the bootstrap quark of model [19, 20] there is a bound state in the gluon channel with mass of the order 0.7 GeV. This bound state should be identified as a constituent gluon. In our consideration we take into account the colour octet state (like $u\bar{d}$) with $J^{PC} = 1^{--}$ and I=1, which determines with the constituent gluon the hybrid state $q\bar{q}g$. The calculated values of mass lowest hybrid mesons are shown in the Table I. The results are in the agreement with the papers [4, 24]. This perturbative calculation corresponds to mixing hybrid and $q\bar{q}q\bar{q}$ states. The absence of experimental data for these states does not to allow to verify the detailed coincidence. In the considered calculation the quark masses $m$ is not fixed. In order to fix anyhow $m$, we assume $m = 570\ MeV$ $(m \geq \frac{1}{4} m_{\hat{a}_2}(2230))$. The model under consideration proceeds from the assumption that the quark interaction forces are the two-component ones. The long-range component is due to the confinement. When the low-lying mesons are considered, the long-range component of the forces is neglected. The creation of ordinary mesons is mainly due to the constituent gluon exchange (Fig. 2(a)). But for the hybrid mesons the long-range forces are important. Namely, the confinement of the $q\bar{q}$ pair with comparatively large energy is actually realized as the production of the new $q\bar{q}$ pair. This means that in the transition $q\bar{q} \to q\bar{q}$ the forces appear which are connected with the process of the Fig. 2(b) type. These box-diagrams can be important in the formation of hadron spectra [25]. We do not see any difficult in taking into account the box-diagrams with the help of the dispersion technique. For the sake of simplicity we restrict ourselves to the introduction of quark mass shift $\Delta$, which are defined by the contributions of the nearest production thresholds of pair mesons $pp$, $ph$, $K\bar{K}$, $Kh$ and so on. We suggest that the parameter $\Delta$ takes into account the confinement potential effectively: $m = m_0 + \Delta$, $m_0 = 0,385$ GeV [19,20], $\Delta = 0,185$ GeV. It changes the behavior pair quark amplitude (1). It allows us to construct the hybrid mesons amplitudes and calculate the mass spectrum hybrid mesons by analogy with the calculation of mass values of the lowest baryons ($J^P = \frac{1}{2}^+, \frac{3}{2}^+$) in the bootstrap quark model [26]. The model in consideration have two parameters: cut-off parameter $\Lambda$ and gluon constant. The subenergy cut-off $\Lambda$ and the vertex function $g$ can be determined by mean of fixing of lowest hybrid meson mass values ($J^{PC} = 0^{++}, 2^{++}$). The vertex functions of various types of the interactions are given in Table II. The mass of the hybrid meson with $J^{PC} = 0^{-+}$ is obtained smaller as compared paper [27 - 29]. It may be determine by them, that the main contribution to state $J^{PC} = 0^{-+}$ is given by the four-particle state. The contribution of hybrid amplitude is only 22 % of the hybrid meson amplitude.

## IV. Conclusion.

In the present paper in the framework of approximate method of solution four-particle relativistic problem the mass spectrum of hybrid mesons, including u, d - quarks, are calculated. The interesting result of this model is the calculation of hybrid meson amplitudes, which contain the contributions of two subamplitudes: four four-quark amplitudes $A_1$ and hybrid amplitudes $A_2$. The contributions of these subamplitudes with different quantum numbers are given in Table I. One can see that the main contributions correspond to the four-quark amplitudes $A_1$. The hybrid contribution corresponds to only less 40 %. We obtained that the small contributions of hybrid give rise to the smaller mass of lowest hybrid meson with $J^{PC} = 0^{-+}$ as compared paper [27 - 29]. The decay width of hybrid mesons can be calculated in the framework this model. The suggested approximate method allows to construct the hybrid meson amplitudes, including heavy quarks Q = s, c, b and calculate the mass spectrum of heavy hybrid mesons.

The two-particle phase space for the equal quark masses is defined as:

$$r_1(s_{ik}, J^{PC}) = \left(a(J^{PC})\frac{s_{ik}}{4m^2} + b(J^{PC})\right)\sqrt{\frac{s_{ik} - 4m^2}{s_{ik}}},$$

$$r_2(s_{ik}) = r_3(s_{ik}) = r_1(s_{ik}, 1^{--}).$$

The vertex functions are shown in Table II. The coefficients $a(J^{PC})$ and $b(J^{PC})$ are given in Table III.

APPENDIX B

The integration contour 1 (Fig. 3) corresponds to the connection $s_{123} < (\sqrt{s_{12}} - m)^2$, the contour 2 is defined by the connection $(\sqrt{s_{12}} - m)^2 < s_{123} < (\sqrt{s_{12}} + m)^2$. The point $s_{123} = (\sqrt{s_{12}} - m)^2$ is not singular, that the round of this point at $s_{123} + ie$ and $s_{123} - ie$ gives identical result. $s_{123} = (\sqrt{s_{12}} + m)^2$ is the singular point, but in our case the integration contour can not pass through this point that the region in consideration is situated below the production threshold of the four particles $s < 16m^2$. The similar situation for the integration over $s_{13}$ in the function $J_3$ is occurred. But the difference consists of the given integration region that is conducted between the complex conjugate points (contour 2 Fig. 3). In Fig. 3, 4b, 5 the dotted lines define the square root cut of the Chew-Mandelstam functions. They correspond to two-particles threshold and also three-particles threshold in Fig. 4(a). The integration contour 1 (Fig. 4(a)) is determined by $s < (\sqrt{s_{12}} - \sqrt{s_{34}})^2$, the contour 2 corresponds to the case $(\sqrt{s_{12}} - \sqrt{s_{34}})^2 < s < (\sqrt{s_{12}} + \sqrt{s_{34}})^2$. $s = (\sqrt{s_{12}} - \sqrt{s_{34}})^2$ is not singular point, that the round of this point at $s + ie$ and $s - ie$ gives identical results. The integration contour 1 (Fig. 4(b)) is determined by region $s < (\sqrt{s_{12}} - \sqrt{s_{34}})^2$ and $s_{134} < (\sqrt{s_{34}} - m)^2$, the integration contour 2 corresponds to $s < (\sqrt{s_{12}} - \sqrt{s_{34}})^2$ and $(\sqrt{s_{34}} - m)^2 \le s_{134} < (\sqrt{s_{34}} + m)^2$. The contour 3 is defined by $(\sqrt{s_{12}} - \sqrt{s_{34}})^2 < s < (\sqrt{s_{12}} + \sqrt{s_{34}})^2$. Here the singular point would be $s_{134} = (\sqrt{s_{34}} + m)^2$. But in our case this point is not achievable, if one has the condition $s < 16m^2$. We have to consider the integration over $s_{24}$ in the function $J_3$. While $s_{124} < s_{12} + 5m^2$ the integration is conducted along the complex axis (the contour 1, Fig. 5). If we come to the point $s_{124} = s_{12} + 5m^2$, that the output into the square root cut of Chew-Mandelstam function (contour 2, Fig. 5) is occurred. In this case the part of the integration contour in nonphysical region is situated and the integration contour along the real axis is conducted. The other part of integration contour corresponds to physical regions. This part of integration contour along the complex axis is conducted. The suggested calculation show that the contribution of the integration over the nonphysical region is small [18].

Table I. Low-lying hybrid meson masses with the isospin I=1 and contributions of four-quark subamplitude $A_1$ and hybrid subamplitude $A_2$ subamplitudes to the hybrid meson amplitude in %.

| $J^{PC}$ | Masses (MeV) | $A_1$ | $A_2$ |
|---|---|---|---|
| $0^{++}$ | $\hat{a}_0$ 1800 (1800) | 71,48 | 28,52 |
| $1^{++}$ | $\hat{a}_1$ 1997 (1940) | 67,29 | 32,71 |
| $2^{++}$ | $\hat{a}_2$ 2230 (2230) | 58,20 | 41,80 |
| $0^{-+}$ | $\hat{p}$ 1364 (1610) | 77,89 | 22,11 |
| $1^{--}$ | $\hat{r}$ 2200 (2020) | 59,84 | 40,16 |

Parameters of model: cut-off parameter $\Lambda = 17,9$; gluon constant $g = 0,373$; effective mass $m = 570$ MeV. The hybrid meson mass values of paper [4, 24] are given in parentheses.

Table II. Vertex functions

| $J^{PC}$ | $G_1^2$ |
|---|---|
| $0^{++}$ | $-8g/3$ |
| $1^{++}$ | $4g/3$ |
| $2^{++}$ | $4g/3$ |
| $0^{-+}$ | $8g/3 - 4g(m_i + m_k)^2 / (3s_{ik})$ |
| $1^{--}$ | $4g/3$ |

The vertex functions $G_1$ correspond to colour singlet states. $G_2^2(s_{ik}) = 2g$ and $G_3^2(s_{ik}) = g$, correspond to the constituent gluon and $u\bar{d}$ state in colour channel $8_c$ with $J^{PC} = 1^{--}$ and I=1 respectively. Here $g$ is the gluon constant. In the present paper the contribution of axial interaction to the state $J^{PC} = 0^{-+}$ is taken into account.

Table III. Coefficient of Chew-Mandelstam functions.

| $J^{PC}$ | $a(J^{PC})$ | $b(J^{PC})$ |
|---|---|---|
| $0^{++}$ | $-1/2$ | $1/2$ |
| $1^{++}$ | $1/2$ | $0$ |
| $2^{++}$ | $3/10$ | $1/5$ |
| $0^{-+}$ | $1/2$ | $0$ |
| $1^{--}$ | $1/3$ | $1/6$ |

Figure captions.

Fig. 1. Graphic representation of the equations for the four-quark subamplitude $A_1(s, s_{12}, s_{123})$ (a) and the hybrid subamplitude $A_2(s, s_{12}, s_{34})$ (b). The bold line corresponds to the constituent gluon contribution.

Fig. 2. Diagram of gluonic exchange defines the short-range component of quark interactions (a) and box-diagram of meson M takes into account the long-range interaction component of the quark forces (b).

Fig. 3. Contours of integration 1, 2 in the complex plane $s_{13}$ for the functions $J_1$, $J_3$.

Fig. 4. Contours of integration 1, 2, 3 in the complex plane $s_{134}$ (a) and $s_{13}$ (b) for the function $J_2$.

Fig. 5. Contours of integration 1, 2 in the complex plane $s_{24}$ for the function $J_3$.

References.

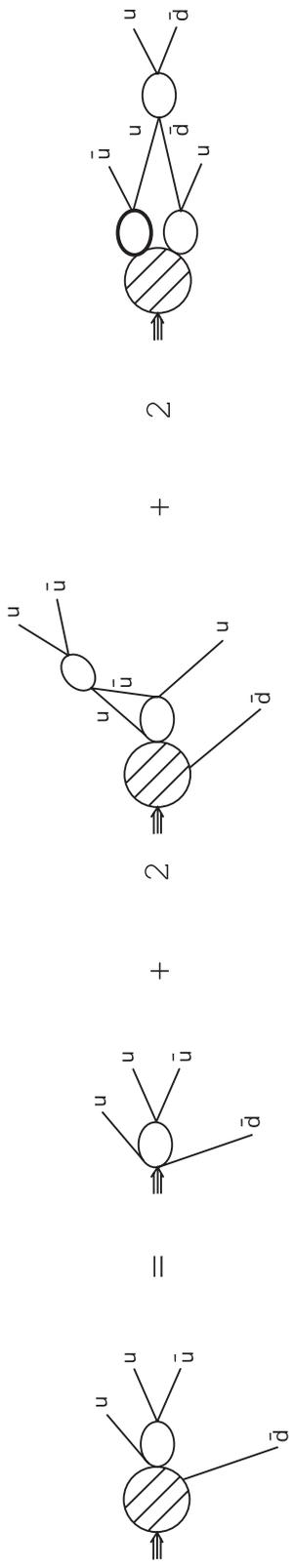

Fig. 1(a)

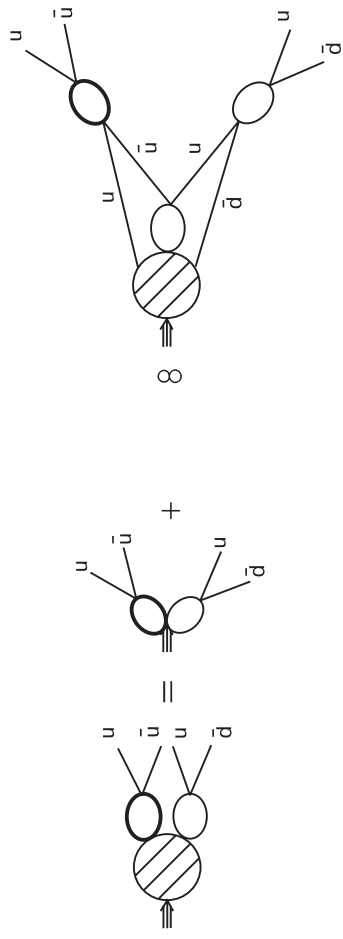

Fig. 1(b)

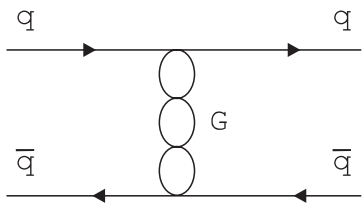

Fig. 2(a)

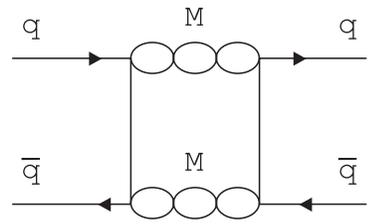

Fig. 2(b)

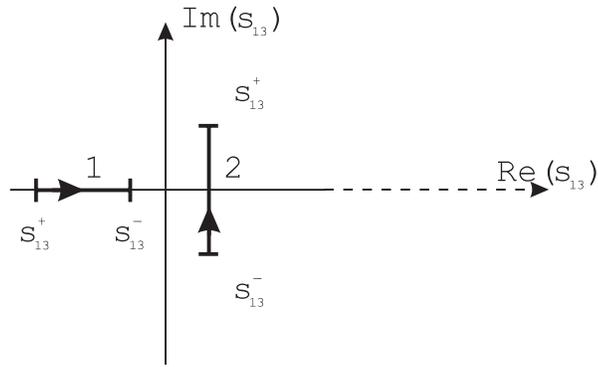

Fig. 3

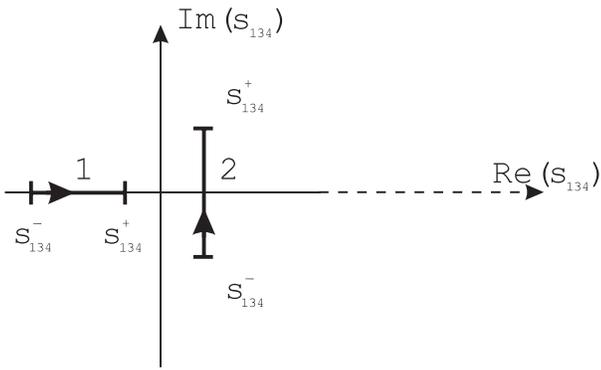

Fig. 4(a)

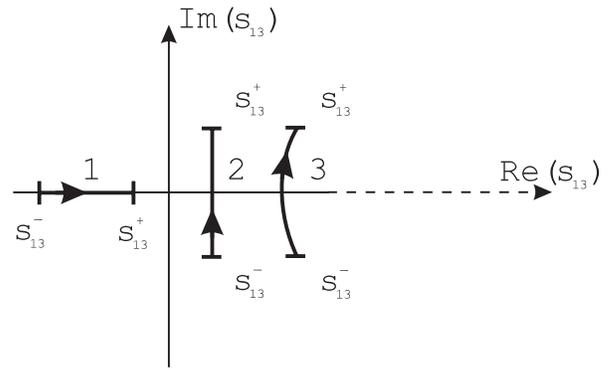

Fig. 4(b)

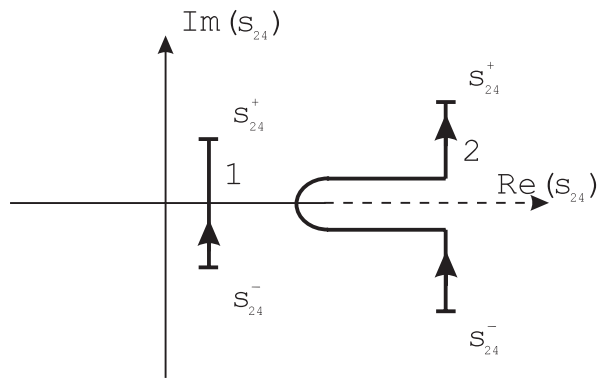

Fig. 5